\documentclass[12pt,a4paper]{article}
\usepackage{graphicx}
\usepackage{amssymb}
\usepackage{amsmath}
\usepackage{bm}
\usepackage{color}
\usepackage{theorem}
\usepackage{subfigure}
\usepackage{amsfonts}
\usepackage{bm}

\usepackage[sort&compress,numbers, merge]{natbib}

\definecolor{Orange}{cmyk}{0,0.61,0.87,0}
\definecolor{JungleGreen}{cmyk}{0.99,0,0.52,0}
\definecolor{OliveGreen}{cmyk}{0.64,0,0.95,0.40}
\definecolor{Brown}{cmyk}{0,0.81,1,0.60}
\definecolor{RoyalBlue}{cmyk}{0.71,0.53,0,0.12}

\newcommand{\be}{\begin{equation}}
\newcommand{\ee}{\end{equation}}
\newcommand{\bea}{\begin{eqnarray}}
\newcommand{\eea}{\end{eqnarray}}
\newcommand{\eq}[1]{Eq.~(\ref{#1})}
\newcommand{\la}{\langle}
\newcommand{\ra}{\rangle}

\usepackage[colorlinks=True, citecolor=blue, linkcolor=blue, urlcolor=black]{hyperref}

\allowdisplaybreaks[1]

\begin{document}

\begin{titlepage}
\begin{center}
\hfill TUM-HEP-1291/20

\vspace{2.0cm}
{\Large\bf  
Anomalous Dimensions of Effective Theories\\
\vspace{.4cm}
from Partial  Waves}

\vspace{1.0cm}
{\small \bf 
Pietro Baratella$^{a}$,
Clara Fernandez$^{b}$,
Benedict von Harling$^{b}$ and
Alex Pomarol$^{b,c}$
}

\vspace{0.7cm}
{\it\footnotesize
${}^a$Technische Universit\"{a}t M\"{u}nchen, Physik-Department, 85748 Garching\\
${}^b$IFAE and BIST, Universitat Aut\`onoma de Barcelona, 08193 Bellaterra, Barcelona\\
${}^c$Departament de F\'isica, Universitat Aut\`onoma de Barcelona, 08193 Bellaterra, Barcelona\\
}

\vspace{0.9cm}
\abstract{On-shell amplitude methods have proven to be extremely efficient for calculating anomalous dimensions.
We further elaborate  on these methods  to show that, 
by the use of an angular momentum decomposition,
the one-loop anomalous dimensions can be reduced to essentially a sum of products of partial waves.
We apply this to the SM EFT, and show how certain classes of anomalous dimensions  have their origin in the same  
partial-wave coefficients.
We also use our result to  obtain a generic formula for the  one-loop anomalous dimensions 
of nonlinear sigma models at any order in the energy expansion, 
and  apply  our method to gravity, where 
  it proves to be very advantageous even in the presence of IR divergencies.}

\end{center}
\end{titlepage}
\setcounter{footnote}{0}

\section{Introduction}

Approaching physical systems by using effective field theories 
has proven to be  quite successful in order to describe  their low-energy  physical properties.
The anomalous dimensions $\gamma_i$ play an important role in this approach, 
since they determine  how the parameters of the theory (the Wilson coefficients)
vary  with the energy scale at which the experiments are carried out. 
Their calculation is therefore important  to extract the relevant properties of the systems,
as shown in many examples in hadron physics, the SM or gravity.

Recently, there has been certain activity to show how
to calculate anomalous dimensions from on-shell methods  \cite{Bern:2019wie,EliasMiro:2020tdv,us,Bern:2020ikv,Jiang:2020mhe},  following the  earlier work of \cite{ArkaniHamed:2008gz,Huang:2012aq,Caron-Huot:2016cwu}.
These methods are equivalent to the ordinary operator approach of effective field theories, since 
amplitudes can be related one-to-one to higher-dimensional operators.
Nevertheless, on-shell  methods have been shown to be  simpler and more efficient,   and also more manageable
 when going beyond the one-loop order (see for example \cite{EliasMiro:2020tdv,Bern:2017puu,Bern:2020ikv}).

In particular, in Ref.~\cite{us} a simple formula was presented for computing anomalous dimensions 
from a product of tree-level amplitudes, integrated over some phase space.
In this paper, 
we will  gain a deeper understanding of the properties of this formula by performing a partial-wave decomposition of the amplitudes.
This will allow us  to effectively capture the information contained in the conservation of angular
momentum (see also \cite{Jiang:2020sdh} for related work). 
As we will see in detail, angular-momentum analysis allows us to reduce the
phase-space integrals in \cite{us} to  a product  of two partial-wave coefficients, $a^J_L a^J_R$,
summed over the angular momenta $J$. The key observation is that,
for contact interactions (in particular, those related to higher-dimensional operators), only  one or a few $J$s contribute to the decomposition, so that the sum over $J$ has a reduced number of  terms.
In other words,  the  non-trivial information necessary to obtain  the anomalous dimension is 
contained in a few partial-wave coefficients. 
 
The traditional arena for using partial-wave analysis is two-to-two scatterings or, equivalently, 4-point amplitudes, where the method is already quite developed \cite{Jacob:1959at}.
In this article, we will restrict ourselves to these amplitudes, as they are  also  the important ones  in  many  theories of interest.
The generalization to higher-point amplitudes can however be  easily obtained  by following the lines of \cite{Jiang:2020sdh}.
The presence of IR divergencies in the renormalization of amplitudes
will also be considered, and we will see that this requires a regularization of the partial-wave coefficients.

We will present  various applications of our results.
First, we will consider the  SM EFT 
and calculate  several anomalous dimensions.  We will see that, in a class of 4-point amplitudes with  equal total helicity,
the partial-wave coefficient which is needed to calculate the one-loop mixings is essentially  the same.
After that, we will see that our approach  also allows us to obtain a generic formula for the anomalous dimensions of
 nonlinear sigma models at any order in the energy expansion.
 Finally, we will show how the calculations in
 gravity  are even simpler than in the previous examples, and definitely much simpler than
 in the ordinary Feynman approach.

\section{Anomalous dimensions  from partial-wave amplitudes}

In this work, we follow the on-shell amplitude approach, and define a theory by its particle content and scattering amplitudes. Higher-point amplitudes are constructed from lower-point ones, with  the ones with
the least points playing the role of building-blocks of the theory.
In low-energy EFfective Theories (EFTs), the set of ``building-block" amplitudes ${\cal A}_i$ is constructed
according to an expansion in $E/\Lambda$, where $\Lambda$ is some UV cut-off of the EFT.
Here, we will be considering EFTs with only massless states, and follow the spinor-helicity notation \cite{Dixon:2013uaa}.
For  scalars, fermions, gauge bosons and  gravitons, the leading term in 
the $E/\Lambda$ expansion is given by
3-point amplitudes  (which can in general be {\it non-local}) defined at complex momenta.
From these, one can build for example all  amplitudes of gauge theories.
At  higher order in $E/\Lambda$,   
the amplitudes ${\cal A}_i$  correspond to  contact interactions,  
which will be  the subject of our interest.
Among them,  we will focus on 3-point and 4-point amplitudes.

The   independent coefficients  of 
the building-block  amplitudes ${\cal A}_i$, which we denote as $C_i$,
 are the ``couplings" of the EFT, which must of course be determined experimentally.
 They are renormalized at the loop level, receiving anomalous dimensions $\gamma_i$.
We want to  show that  these anomalous dimensions at the one-loop level can easily be determined from products of   partial-wave amplitudes.

\subsection{One-loop anomalous dimensions via on-shell amplitudes}

Let us start by considering the case in which there are no IR divergencies.
It was shown in \cite{us} that, in this situation, the one-loop anomalous dimension of an amplitude $\gamma_i$
can be written as a product of tree-level amplitudes under a phase-space integration.
Defining $\gamma_i=d C_i/d\ln\mu$, we find for the particular case of 4-point amplitudes that \cite{us}
\be
\label{MF}
\gamma_i \frac{{\cal A}_i (1,2,3,4)}{C_i}
=-\frac{1}{4\pi^3}
\int d{\rm LIPS}\sum_{1',2'}\sigma_{1'2'} 
\Big[ {\cal A}_L(1,2,{ \bar 2'},{ \bar 1'}) {\cal A}_R(1',2',3,4)\Big]
+(2\leftrightarrow 3)
+(2\leftrightarrow 4)\,.
\ee
Here ${\cal A}_{L,R}$ are  4-point tree-level amplitudes (with all states incoming),
which are constructed using the building-block amplitudes ${\cal A}_j$.
Their exponent in  $1/\Lambda$, which we denote as $w$, {\it i.e.}~${\cal A}\propto 1/\Lambda^w$, must satisfy
\be
\label{wsr}
w_i=w_L+w_R\,.
\ee
The three terms in \eq{MF}  arise from 2-cuts of the one-loop amplitude in respectively the $s$-, $t$- and $u$-channel.\footnote{The external particles of the $t$-channel term are ordered as 1,3,2,4. If this
reordering (starting from 1,2,3,4) implies an odd number of fermion exchanges, the $t$-{channel} term of \eq{MF}
has a minus sign. Similarly for the $u$-channel.} 
The sum  $\sum_{1',2'}$ in \eq{MF} is taken over amplitudes ${\cal A}_{L,R}$ with all possible internal states $1',2'$ (and over quantum numbers such as color or flavor).
The bar over a state (e.g.~$\bar 1$) indicates that it  
carries  opposite-sign momentum, helicity and all other quantum numbers with respect to the state without a bar (e.g.~$1$). When writing down amplitudes in the spinor-helicity formalism (as we do here),
one needs a prescription for $\left|-p\right\rangle$ and $\left|-p\right]$. Here we take
$\left|-p\right\rangle=i|p\ra$ and $\left|-p\right]=i|p]$ \cite{us}. In this convention,
the factor $\sigma_{1'2'}$ is defined by $\sigma_{i_1i_2}\equiv(-i)^{F_{i_1i_2}}$, where $F_{i_1i_2}$
counts the number of fermions in the list $\{i_1i_2\}$.\footnote{Notice that in \eq{MF}  we put the states with negative momentum,  $\bar 1'$ and $\bar 2'$,  in  ${\cal A}_L$, while in the last version of \cite{us} we put them in ${\cal A}_R$. For this reason the factor $\sigma_{i_1i_2}$ in \cite{us} was defined as $\sigma_{i_1i_2}=i^{F_{i_1i_2}}$.}
A factor 1/2 must also be  included in \eq{MF} when the internal particles are indistinguishable.

Before proceeding, let us point out simple generalizations of \eq{MF} which we will use in  the examples
of Sec.~\ref{exsec}. 
Firstly, it is quite common to have more than one independent amplitude ${\cal A}_i$ with the same 4 external states.
In this case, we have a sum over the corresponding indices $i$ on the LHS of \eq{MF}.
Secondly, the ${\cal A}_i$ in \eq{MF} does not necessarily have to be a building-block amplitude. Instead, it can be
a 4-point amplitude made from 3-point building-blocks. In this case, we will denote it as $\widehat {\cal A}_i$.
We will see an example of this in the context of gravity, where we will study
the renormalization of the 3-point amplitude made of equal-helicity gravitons, ${\cal A}_{R^3}$,
from  the renormalization of a 4-point amplitude containing it, which we call $\widehat {\cal A}_{R^3}$.

As a next step, we will
perform a decomposition into the angular momenta $J$ of the internal state,
the particle pair $1',2'$ (and, similarly,
decompose it according to its other quantum numbers such as isospin).
For a given $J$ of the $1',2'$-system,
angular momentum conservation implies that also the external states of ${\cal A}_{L,R}$ 
must be in the same $J$, and the same for ${\cal A}_i$. 

The interesting point is that, for contact interactions like
${\cal A}_i$ with $w_i>0$, only a few $J$-channels contribute, as we will show with some examples.
This means that the anomalous dimension in \eq{MF} is determined by only a few $J$-channels,
 simplifying then its computation. 
The purpose of the following analysis is to make these statements quantitative,
 finding out  explicitly  what  angular momenta $J$ mediate the anomalous dimensions.

\subsection{ Partial-wave decomposition}

In order to perform the angular-momentum decomposition of an amplitude {${\cal A}(1_{h_1},2_{h_2},3_{h_3},4_{h_4})$, with $h_i$ denoting the helicity of particle $i$}, it is convenient to specify the pair of  incoming and outgoing states.
Let us for concreteness consider the $s$-channel,  {\it i.e.}~$1_{h_1},2_{h_2}\to 3_{-h_3},4_{-h_4}$.
{We define}
\be
{\cal A}(1_{h_1},2_{h_2}\to 3_{-h_3},4_{-h_4})\equiv \sigma_{34}\,{\cal A}(1_{h_1},2_{h_2},-4_{h_4},-3_{h_3})\,,
\ee
{where the factor $\sigma_{34}$ when crossing fermions {arises in} our
convention where $\left|-p\right\rangle=i|p\ra$ and $\left|-p\right]=i|p]$}.
Notice that, in terms of the `bar' notation
previously introduced, one has for example $\overline{(1_{h_1})}=-1_{-h_1}$.
By going  to the  center-of-momentum frame 
and aligning the $z$-axis with the direction of particles 1 and 2 (with $\vec p_1$ pointing downwards),
we can parametrize the direction of the outgoing  particle {$4$} by the polar  coordinates $(\theta,\phi)$. 
The amplitude can then be written as a function of the two polar angles and the Mandelstam variable $s$:
 \be\label{comA}
{\cal A}(1_{h_1},2_{h_2}\to 3_{-h_3},4_{-h_4})={\cal A}(s,\theta,\phi)\,.
 \ee
When the amplitude is given in 
spinor-helicity notation, this frame corresponds to taking
{(notice that here $p_1+p_2=p_3+p_4$)}
\be
\begin{array}{l}
 |3\ra=  c_{\theta/2}| 1\ra -s_{\theta/2} e^{-i\phi}|2\ra\\
|4\ra=  s_{\theta/2} e^{i\phi}|1\ra+c_{\theta/2}|2\ra
\end{array} , \quad \ \
\begin{array}{l}
|3]=c_{\theta/2}| 1] -s_{\theta/2} e^{i\phi}|2]\\
|4]=s_{\theta/2} e^{-i\phi}|1]+c_{\theta/2}|2]
\end{array},
\label{34}
\ee
where $s_{\theta/2}\equiv\sin\theta/2$ and $c_{\theta/2}\equiv\cos\theta/2$.
Furthermore, we have  $\la 12\ra=[ 21]=\sqrt{s}$,
while the Mandelstam variables $t =-\la 13\ra [ 31]$ and $u = -\la 14\ra [ 41]$ become 
\be
t=-s\,\frac{1-c_\theta}{2}\ ,\quad \quad  u=-s\,\frac{1+c_\theta}{2}\,.
\label{tu}
\ee
To exploit the information contained in the conservation of angular momentum,
we can now perform a partial-wave decomposition of the initial and final states of
${\cal A}$ {in \eq{comA}} (on a basis having definite angular momentum, quantized along the $z$-axis).
This gives \cite{Jacob:1959at} (see details in Appendix~\ref{AppB})
\be
{\cal A}(s,\theta,\phi)
=e^{i\phi(h_{12}-h_{43})}\left(\frac{\sqrt{s}}{\Lambda}\right)^w\sum_J n_J\, d_{h_{12}h_{43}}^J(\theta)\,a^J ,
\label{pwd}
\ee
where $n_J=2J+1$,  $h_{12}=h_1-h_2$ (and similarly for $h_{43}$), and where $d_{hh'}^J(\theta)$ are the Wigner $d$-functions. 
Notice that we have factored out  the dependence of the amplitude on $\sqrt{s}/\Lambda$.
The partial-wave expansion, \eq{pwd}, can be inverted
to give the coefficients $a^J$ for the amplitude:
\be
\label{aj_formula}
a^J=\frac{1}{2}\left(\frac{\sqrt{s}}{\Lambda}\right)^{-w} \hspace{-.1cm}\int_0^\pi d\theta\,  s_\theta \,d_{h_{12}h_{43}}^J(\theta) \, {\cal A}(s,\theta,\phi=0)\,,
\ee
where we have used the orthogonality of the Wigner $d$-functions (see \eq{ortho} in Appendix~\ref{AppB}).
The above derivation implicitly assumes the existence of well-defined coefficients $a^J$,
which is not always guaranteed, as we will see in Sec.~\ref{soft}.

In the same way as for the $s$-channel, we can perform a partial-wave decomposition for the 
 $t$-channel $1,3\to{2,4}$ and the $u$-channel $1,4\to{3,2}$.
 This leads to  the same expression  as \eq{pwd}, with the replacements
$(2\leftrightarrow 3)$ and $(2\leftrightarrow 4)$ respectively. In this case, the polar angles
give the direction of the outgoing pairs {$2, 4$} and {$3,2$} respectively.
The coefficients $a^J$, either derived in the $s$-, $t$- or $u$-channel,
completely characterize the amplitude ${\cal A}(1,2,3,4)$. 
Depending on the problem under consideration, we can use one or another. 
We can also write \eq{pwd}
in a  manifestly Lorentz-invariant form by inverting \eq{tu}, which leads to
$c_\theta=(t-u)/s$, while in the $t$- and $u$-channel we have $c_\theta=(s-u)/t$  and $c_\theta=(t-s)/u$, respectively.

{Let us now consider the phase-space integrations in \eq{MF}, 
and apply a partial-wave decomposition to  ${\cal A}_{L}$ 
and ${\cal A}_{R}$  with respect to the $J$ of the two external states.
This corresponds to  decomposing ${\cal A}_{L,R}$   in  the first  term of \eq{MF} 
in the $s$-channel, {specifically $1,2\to 1' ,2'$ and $1',2' \to \bar 4,\bar 3$}.
Let us show in detail how this proceeds.
We denote the polar angles which give the $1',2'$-direction as $(\theta',\phi')$, and those for the
{$\bar 4,\bar 3$}-direction as $(\theta,\phi)$.}
The phase-space integral for $1',2'$  in \eq{MF}  then reduces to an angular integration over the primed polar coordinates, that is $\int d{\rm LIPS}=\int d\theta' s_{\theta'} d\phi'/8$. Using \eq{general} of
Appendix~\ref{AppB}, we find
\bea
\label{aLaR}
&&\int d{\rm LIPS}~\sigma_{1'2'}{\cal A}_L(1,2,\bar 2',\bar 1') {\cal A}_R(1',2',3,4)
=  \int d{\rm LIPS}~{\cal A}_L(1,2\to 1',2')\, \sigma_{34}^{-1}{\cal A}_R(1',2'\to \bar 4,\bar 3)\nonumber\\
&&=\frac{1}{8}\,\sigma_{34}^{-1}\left(\frac{\sqrt{s}}{\Lambda}\right)^{w_L+w_R}\hspace{-.1cm}\int_0^\pi d\theta' s_{\theta'} \int_0^{2\pi} d\phi' \, \sum_{J'}n_{J'}\,e^{i\phi'(h_{12}-h'_{12})}d_{h_{12}h'_{12}}^{J'}(\theta')\, a_L^{J'}\nonumber\\
&&\hskip3.4cm
\times \sum_{JM}n_J\,e^{i\phi(M-h_{34})}
d_{Mh_{34}}^J(\theta)e^{-i\phi'(M-h'_{12})}d_{Mh'_{12}}^J(\theta')\,a_R^J\nonumber\\
&&=\frac{\pi}{2}\,\sigma_{34}^{-1}\,e^{i\phi(h_{12}-h_{34})}\left(\frac{\sqrt{s}}{\Lambda}\right)^{w_L+w_R}\sum_J n_J \, d^J_{h_{12}h_{34}}(\theta) \,a_L^J a_R^J\,,
\eea
where in the last step we have performed a trivial $d\phi'$ integration and used the orthogonality
condition of the Wigner $d$-functions (\eq{ortho} of Appendix~\ref{AppB}).
{Similarly, we can proceed as   above for the second and third term of \eq{MF}, which  in this case must be decomposed
in the $t$- and $u$-channel respectively. 
Inserting the  result in \eq{MF}, we obtain}
\be
\gamma_i\frac{{\cal A}_i}{C_i}
=-\frac{1}{8\pi^2}\,\sigma_{34}^{-1}\,
 e^{i\phi(h_{12}-h_{34})}\,\left(\frac{\sqrt{s}}{\Lambda}\right)^{w_L+w_R}
\sum_J n_J\, d^J_{h_{12}h_{34}}(\theta)\sum_{1'2'} a_L^J a_R^J
+t\text{-channel}+u\text{-channel}\,.
\label{NMF}
\ee
The sum over $J$ can in principle be infinite.
Nevertheless,  4-point amplitudes
${\cal A}_i$  with $w>0$ (which are contact interactions)
have  a finite number of partial waves
(the incoming states can only be in a few $J$s), 
implying that also the RHS of \eq{NMF} consists of a finite number of nonzero terms.
In most of the cases this is true because only a few coefficients $a^J_L$ or $a^J_R$ are nonzero,
which makes \eq{NMF}  a very simple formula to  calculate   anomalous dimensions.
 But this is not always the case.
Indeed, it is still possible that an infinite series in $J$ appears on the RHS of \eq{NMF}.
When this happens, a non-trivial cancellation between the 3 different channels
 takes place,  such that only
a finite number of  $J$ contributions remain on the RHS as expected. 
This occurs however only in a few circumstances, as 
we will discuss at the end of  Sec.~\ref{SMex}.

\eq{NMF} simplifies enormously in certain cases.
For example, if there are only  contributions from one channel, say the $s$-channel, we can expand
the amplitude ${\cal A}_i$ on the LHS of \eq{NMF} into partial waves in the same channel. From this we get
\be
\gamma_{i}\frac{a_{i}^J}{C_i} =-\frac{1}{8\pi^2}\sum_{1'2'} \, a_L^Ja_R^J\,.
\label{NMFS}
\ee
When several ${\cal A}_i$ contribute on the LHS of \eq{NMF},
we instead have a sum over the corresponding indices $i$ on the LHS of \eq{NMFS}, leading to
a system of equations (obtained by projecting on the corresponding quantum numbers of the  ${\cal A}_i$)
that must be solved to obtain the individual $\gamma_i$.
We will see an example in Sec.~\ref{pions}.
{We further note that $\sigma_{34}^{-1}$ drops out
in \eq{NMFS}.}

A similar  simplification arises when the contribution from each channel is independently proportional to 
a single ${\cal A}_i$.\footnote{We stress that this does not always happen. In general, the RHS of  \eq{NMF} becomes proportional to ${\cal A}_i$ only after adding the contributions from the 3 channels.}
 This occurs, for example,  when each channel gives a contribution that is  parametrically independent from the others
 (because each depends on different couplings). In this case,  we can write
 \be
 \gamma_i= \gamma^s_i+ \gamma^t_i+ \gamma^u_i \,,
\label{total}
 \ee
 where $\gamma_i^{s,t,u}$ is the contribution from the $s$-,$t$-,$u$-channel respectively.
 For  the $s$-channel, this contribution is given by
 \be 
\gamma^s_i\frac{{\cal A}_i}{C_i}
=-\frac{1}{8\pi^2} \,\sigma_{34}^{-1}\,e^{i\phi(h_{12}-h_{34})}\left(\frac{\sqrt{s}}{\Lambda}\right)^{w_L+w_R}
\sum_J n_J\,  d^J_{h_{12}h_{34}}(\theta)
\sum_{1'2'}\,a_L^J a_R^J\,.
\label{scont}
\ee
After expanding  ${\cal A}_i$  into partial waves in the  $s$-channel,  \eq{scont} leads to
\be 
 \gamma^s_{i}=-\frac{C_i}{8\pi^2}\sum_{1'2'} \frac{a_L^Ja_R^J}{a_{i}^J}\,.
\label{schannel}
\ee
 In a similar way, we can proceed for  the contributions from the other channels.

Although  these particular cases could  look very special, they are actually quite common. 
For example, in the SM EFT most  of the renormalizations of  4-point amplitudes  at $O(E^2/\Lambda^2)$
can only proceed through one partial wave in one channel,
as we will see in the examples of Sec.~\ref{SMex}.
On the other hand, for EFTs of  Goldstone bosons or gravitons,
we will have to add the  
three channels together as in \eq{NMF}. 
Nevertheless,  we will  only need to calculate the $s$-channel contribution,
since the other channels are determined by crossing symmetry.

\subsection{IR divergencies}\label{soft}

There are certain cases in which the $d$LIPS integral in \eq{MF} is divergent and must be regulated.
This is due to the singular behavior of either $\mathcal{A}_{L}$ or $\mathcal{A}_{R}$
in the limit $\theta\to 0$ or $\theta\to \pi$ (or both), with the amplitude going respectively like
$\mathcal{A}\sim s_{\theta/2}^{-2}~{\rm or~}c_{\theta/2}^{-2}~{\rm (or~}s_{\theta}^{-2})$. We
will focus for the moment on the first case, and later on will comment on the last case.
The second case, with singularities only for $\theta\to \pi$,  can always be excluded by a reordering of legs.

 These  singularities are intimately related to soft IR divergencies of the loop
integral, as explained in Appendix~\ref{appA}. There, we show that in the presence of  $\theta\to 0$ soft IR divergencies, 
\eq{MF} must be corrected by adding the following expression:
\begin{multline}
\label{IR}
\Delta \gamma_i \, \frac{{\cal A}_i(1^{a},2^{b},3^{c},4^{d})}{C_i} \, = \, -\frac{1}{4\pi^3} \left[({\bm T}_{\rm soft}^{12})_{\hat{a}\,\hat{b}}^{a\,b}\,
{\cal A}_R(1^{\hat{a}},2^{\hat{b}},3^{c},4^{d})\,\int d{\rm LIPS}_{12} \,\frac{1}{s_{\theta'\!/2}^2} \right. \\
 \left. + \,  ({\bm T}_{\rm soft}^{34})_{\hat{c}\,\hat{d}}^{c\,d}\,
{\cal A}_L(1^{{a}},2^{{b}},3^{\hat{c}},4^{\hat d})\,\int d{\rm LIPS}_{34} \,\frac{1}{s_{\theta'\!/2}^2}\right]+(2\leftrightarrow 3)
+(2\leftrightarrow 4)\,,
\end{multline}
where  the $d$LIPS$_{ij}$  integrals are over the $i'j'$-state phase space.
\eq{IR} acts as a regulator for the   small-angle divergencies contained in \eq{MF}.
The soft operator ${\bm T}_{\rm soft}^{ij}$ can in general act on color/flavor
indices $a,b,c,d$ and contains
couplings and powers of $s_{ij}=(p_i+p_j)^2$.
In the simplest cases, like QED or gravity, it is diagonal in color/flavor, {\it i.e.}~${\bm T}_{\rm soft}\propto \delta_{a\hat{a}}\,\delta_{b\hat{b}}$ (see
Appendix~\ref{appA} for explicit expressions and the example at the end of Sec.~\ref{SMex}).

When an amplitude  features small-angle singularities,
its  coefficients $a^{J}$ are logarithmically divergent.\footnote{In particular, we have
$a^J\sim \lim_{\epsilon\to 0}\int_\epsilon^\pi {d\theta}/{\theta}\sim \ln\epsilon$.}
Nevertheless, there is a natural generalization of \eq{aj_formula} 
which consists in defining
``regularized" partial-wave coefficients  as
\be
a^{J}|_{\rm reg}=\frac{1}{2}\left(\frac{\sqrt{s}}{\Lambda}\right)^{-w}\int^\pi_0 d\theta\, s_\theta 
\left(d^J_{h_{12}h_{43}}(\theta){\cal A}(1_{h_1}^{a},2_{h_2}^{b}\to {3}_{-h_3}^{c},{4}_{-h_4}^{d})|_{\theta,\phi=0} +\frac{\left({\bm T}_{\rm soft}\right)^{a\,b}_{c\,d}}{s^2_{\theta/2}} \right)\,.
\label{aIRsafe}
\ee
In terms of these coefficients, the anomalous dimensions can still be written as in
\eq{NMF}.
\eq{aIRsafe} can be 
inferred in the following way. 
Let us assume that one of the amplitudes,
either $\mathcal{A}_L$ or $\mathcal{A}_R$, has divergent coefficients $a^J$,
while for the other one all partial-wave coefficients are finite.
Notice that the latter assumption is always fulfilled for contact interactions like the EFT amplitudes ${\cal A}_i$ described above.
For concreteness, we take $\mathcal{A}_R$ to have all coefficients finite.
In this case, the $d$LIPS integrals
of  \eq{MF}, together with those of \eq{IR}, can be written as
\begin{align}
& \int d{\rm LIPS} \left(\sigma_{1'2'}\,{\cal A}_L(1,2,\bar 2',\bar 1') {\cal A}_R(1',2',3,4)
+\frac{{\bm T}_{\rm soft}}{s^2_{\theta' \!/2}}{\cal A}_R(1,2,3,4)\right)\nonumber \\
&  \ \  =\frac{\pi}{2}\,\sigma_{34}^{-1}\,e^{i\phi(h_{12}-h_{34})} \!
\left(\frac{\sqrt{s}}{\Lambda}\right)^{w_R} \! 
\sum_J n_J\, d_{h_{12}h_{34}}^J(\theta) \, a_R^J~\frac{1}{2} \! \int_0^\pi \! d\theta' s_{\theta'} \! \left(d^J_{h_{12}h'_{12}}(\theta'){\cal A}_{L}\smash{(s,\theta',0)} +\frac{{\bm T}_{\rm soft}}{s^2_{\theta'\!/2}} \right)\nonumber\\
& \ \ =\frac{\pi}{2}\,\sigma_{34}^{-1}\,e^{i\phi(h_{12}-h_{34})}\left(\frac{\sqrt{s}}{\Lambda}\right)^{w_R+w_L}\sum_J  n_J\, d_{h_{12}h_{34}}^J(\theta) \, a_R^J\, a_L^J|_{\rm reg}\,,
\label{aIRproof}
\end{align}
where we have left color/flavor indices implicit.
\eq{aIRproof} gives  the generalization of \eq{aLaR}  when there are soft IR divergencies. 
In other words, when the coefficients $a^J$ of an amplitude are ill-defined, signaling the presence of soft IR divergencies,
we can still use \eq{NMF}, but with the regularized coefficients given in \eq{aIRsafe}.
We will see  examples in the next section.

We comment now on the possibility that amplitudes
have singularities for both $\theta \to 0$ and $\theta\to \pi$,
{\it i.e.} ${\cal A}\sim 1/s^2_\theta$,
which happens when the particles $1'$ and $2'$ are identical.
In this case, 
the divergent integral $\int s^{-2}_{\theta'\!/2}$   in \eq{aIRsafe}
must  be replaced by $2\int s^{-2}_{\theta'}$ to make the integrand
 well-behaved (see Appendix~\ref{appA}).
Moreover, \eq{aIRsafe} has to be evaluated only for even $J$, since the coefficients $a^J$  for odd $J$
are zero\footnote{This is because 
$\mathcal{A}$ is even in $[0,\pi]$,
{\it i.e.}~$\mathcal{A}(\theta)=\mathcal{A}(\pi-\theta)$ while, for odd $J$, $s_\theta d^J_{00}(\theta)$ is odd.
Nevertheless, since  ${\cal A}$ has singularities, the integral has to be performed carefully,
by taking $\lim_{\epsilon\to 0}\int_\epsilon^{\pi-\epsilon}
d\theta \, s_\theta \, {\cal A}(\theta)d^J_{00}(\theta)$.
}
and do not need a subtraction.

When also collinear IR divergencies  are present in the  one-loop amplitude, one must add
the extra contribution (see Appendix~\ref{appA})
\be
\Delta \gamma_i \, \frac{{\cal A}_i(1,2,3,4)}{C_i}=
\gamma_{\rm coll}\,{\cal A}_i (1,2,3,4)\,.
\label{coll}
\ee
This is always diagonal in amplitude space and depends only on
the external legs,   so that $\gamma_{\rm coll}=\sum_{n=1}^4 \gamma_{\rm coll}^{(n)}$,
where the $\gamma_{\rm coll}^{(n)}$ are given for example in
\cite{EliasMiro:2020tdv,Jiang:2020mhe}.

\section{Applications}
\label{exsec}

\subsection{The Standard Model EFT}
\label{SMex}

In Refs.~\cite{us,EliasMiro:2020tdv,Bern:2020ikv,Jiang:2020mhe},
several examples of the calculation of the anomalous dimensions  in the SM EFT
via on-shell  amplitude methods were presented.\footnote{Amplitude methods have also been applied  at tree-level in the SM EFT  \cite{Shadmi:2018xan}.}
In particular, in \cite{us} the renormalization of the electron dipole operator was provided,
and a similarity was shown in the contributions to the anomalous dimension coming  from very different  Feynman diagrams. 
As we will see, the angular-momentum analysis clarifies the origin of this similarity.

As a first example, let us consider the one-loop mixing between  4-point  amplitudes 
of total helicity $h=-2$ at $O(E^2/\Lambda^2)$:
\be
\begin{aligned}
	{\cal A}_{WHle}(1_{e},2_{l_j},3_{W^a_-},4_{H^\dagger_i})&
	=\frac{C_{WHle}}{\Lambda^2}\la 31\ra\la 32\ra (T^a)_{ij}\,,\\
	{\cal A}_{eluq,0}(1_{e},2_{l_i},3_{u},4_{q_j})&=
	\frac{C_{eluq,0}}{\Lambda^2}\la 12\ra\la 34\ra \, \epsilon_{ij} \,,\\
	{\cal A}_{eluq,1}(1_{e},2_{l_i},3_{u},4_{q_j})&=
	\frac{C_{eluq,1}}{\Lambda^2}\frac{1}{2}\left(\la 23\ra\la 41\ra+\la 13\ra\la 42\ra\right)\epsilon_{ij}  \,,\\
	{\cal A}_{W^2H^2}(1_{W_-^a},2_{H_j},3_{W_-^b},4_{H^\dagger_i})&=
	\frac{C_{W^2H^2}}{\Lambda^2}\la 13\ra^2\delta_{ab}\delta_{ij}\,.
	\label{SMEFT}
\end{aligned}
\ee
Notice that  we have chosen  ${\cal A}_{eluq,0}$ (${\cal A}_{eluq,1}$) to be
antisymmetric (symmetric) with respect to $1\leftrightarrow 2$. As we will see, its only non-vanishing partial-wave component then has $J=0$ ($J=1$).\footnote{In \cite{us}, the basis was  ${\cal A}_{lequ}\sim\la 12\ra\la 34\ra$ 
and ${\cal A}_{luqe}\sim\la 23\ra\la 41\ra$.}
{In \eq{SMEFT} there are 3 types of amplitudes which differ by the number of fermions involved, $n_F=0,2,4$.
We will study the one-loop mixing between these 3 types.}
Although from the ordinary Feynman approach the  anomalous dimensions of \eq{SMEFT}
arise from very different  diagrams, from 
  on-shell methods we can easily understand that the different mixings    are in fact related.
Indeed,  from \eq{MF} one can realize that the mixings between  amplitudes {with different $n_F$} in \eq{SMEFT} 
can only proceed through the same type of SM amplitude. This is given by  \cite{us}
\be
{\cal A}_{\rm SM}
(1_{\bar \psi_R},2_{\bar \psi_{L_i}},
3_{W^a_-},4_{H_j})
=y_\psi \, g_2 \, (T^a)_{ij}\frac{\la 13\ra \la 43\ra}{\la 14\ra\la 12\ra}\,,
\label{su2}
\ee
or its complex conjugate,
where  $\psi_L\left(\psi_R\right)$  refers to the SM $SU(2)_L$-doublet (singlet) lepton, or to the up-type quark upon the replacement $H_j\to H^\dagger_{j}$ and  $(T^a)_{ij}\to (T^a)_{ij'}\epsilon_{j'j}$.

The partial-wave decomposition can tell us even more.
First, we notice that the mixing between  amplitudes {with different $n_F$} in \eq{SMEFT} proceeds 
only through the  $s$-channel (no product of amplitudes can be found in the $t$- and $u$-channel which can generate these mixings). {The only exception is the renormalization of ${\cal A}_{W^2H^2}$ by ${\cal A}_{WHle}$, which occurs through both the $s$- and the $t$-channel, but these are trivially related by the crossing of the external $W$ bosons.} In 
{the $s$-}channel,  the partial-wave coefficients of  \eq{su2}  
are given (using \eq{34}) by
\be
a^{J=0}_{\rm SM}=0\, ,\ \ \ \ \ a^{J\geq 1}_{\rm SM}={ y_\psi}g_2 (T^a)_{ij}\frac{1}{2}\int_0^\pi d\theta s_\theta \,d_{0,1}^J(\theta)\frac{s_{\theta/2}}{c_{\theta/2}}=\frac{{ y_\psi}g_2 (T^a)_{ij}}{ \sqrt{J(J+1)}}\,,
\ee
while for the  amplitudes of \eq{SMEFT} we report them in  Table~\ref{tableSMEFT}.
Notice that ${\cal A}_{eluq,0}$, having only a $J=0$ partial-wave component,
cannot mix with the rest of the amplitudes which have only $J=1$ components
(this angular  momentum selection rule was already pointed out in Refs.~\cite{Jiang:2020sdh,us}).
The other  amplitudes of \eq{SMEFT} can mix among themselves, but always
through the $J=1$ partial wave.  Therefore all mixings are proportional to $a^{J=1}_{\rm SM}$.

We can explicitly calculate these mixing using \eq{schannel}. We obtain
\begin{equation}
\begin{pmatrix}
\gamma_{WHle}\ C^{-1}_{WHle} a^{1}_{WHle}\\ 
\gamma_{eluq,1}\ C^{-1}_{eluq,1}a^{1}_{eluq,1}\\ 
\gamma_{W^2H^2}\ C^{-1}_{W^2H^2}a^{1}_{W^2H^2}
\end{pmatrix}=
-\frac{\widetilde a^{J=1}_{\rm SM}}{8\pi^2}
\begin{pmatrix}
\times & -N_c y_u & y_e\\ 
-y_u	& \times & 0\\ 
y_e & 0 & \times
\end{pmatrix}
\begin{pmatrix}
a^1_{WHle}\\ 
a^1_{eluq,1}\\ 
a^1_{W^2H^2}
\end{pmatrix}{+ \ \text{crossing}} \,,
\label{mixing}
\end{equation}
where we have defined $a^{J}_{\rm SM}=y_\psi \widetilde a^{J}_{\rm SM}$, and  omitted  the diagonal entries as they  correspond to self-renormalizations which we do not consider here. {The crossing accounts for the renormalization of ${\cal A}_{W^2H^2}$ by ${\cal A}_{WHle}$ in the $t$-channel. This can be easily obtained by just  interchanging 
$W^a \leftrightarrow W^b$ as the kinematics turns out to be invariant under this crossing.}
Therefore, in \eq{mixing}, the matrix in the RHS is trivially determined
by  color factors, signs due to  fermion permutations and  different Yukawa couplings.
The non-trivial part of the one-loop calculation has gone  into 
the product of the coefficients $a^{1}$ of the amplitudes of \eq{SMEFT} with that of  the SM amplitude in \eq{su2}.
By plugging the values of  Table~\ref{tableSMEFT} into \eq{mixing}, we obtain 
\vspace{.1cm}
\begin{equation}
\begin{pmatrix}
\gamma_{WHle}\\ 
\gamma_{eluq,1}\\ 
\gamma_{W^2H^2}
\end{pmatrix}=
\frac{g_2}{16\pi^2}
\begin{pmatrix}
\times & N_cy_u & -2y_e\\ 
\frac{3}{2}y_u	& \times & 0\\ 
-\frac{1}{2}y_e & 0 & \times
\end{pmatrix}
\begin{pmatrix}
C_{WHle}\\ 
C_{eluq,1}\\ 
C_{W^2H^2}
\end{pmatrix}. 
\end{equation}
{This property of all one-loop anomalous dimensions 
being proportional to the same coefficient $a^J_{\rm SM}$
occurs at $O(E^2/\Lambda^2$)
 for  mixings between
4-point amplitudes with different number of fermions $n_F$ 
and  same total helicity $h$. 
Here, we have shown it for  $h=-2$, but 
the same is  true for 4-point amplitudes with $h=0$.
In this case, the mixings between amplitudes with different $n_F$ are proportional to the $J=1$ partial wave of the SM $\psi\bar\psi HH^\dagger$ amplitude.}

\begin{table}\centering
	\begin{tabular}{||c||c|c|c||} 
		\hline
		& $J=0$ & $J=1$ & \\
		\hline\hline
		${\cal A}_{\rm SM}
		(1_{\bar \psi_R},2_{\bar \psi_{L i}},
		3_{W^a_-},4_{H_j})$  &  $0$ & $\frac{1}{\sqrt{2}}$ & $\times\, y_\psi  g_2\, (T^a)_{ij}$\\
		\hline
		${\cal A}^{I=0}_{\rm SM}(1_{H},2_{H^\dagger},3_{H^\dagger},4_{H})$  & $-\frac{3}{8} $ & $-\frac{3}{2} $ & $\times\, g_2^2$ \\
		\hline \hline
		${\cal A}_{WHle}(1_{e},2_{l_j},3_{W^a_-},4_{H^\dagger_i})$  & $0$ & $\frac{1}{3\sqrt{2}}$ &  $\times\, C_{WHle}(T^a)_{ij}$\\ 
		\hline
		${\cal A}_{eluq,0}(1_{e},2_{l_i},3_{u},4_{q_j})$  & 1 & $0$ & $\times\,C_{eluq,0}\, \epsilon_{ij}$\\ 
		\hline
		${\cal A}_{eluq,1}(1_{e},2_{l_i},3_{u},4_{q_j})$  & $0$ & $\frac{1}{6}$  & $\times\,C_{eluq,1}\, \epsilon_{ij}$ \\ 
		\hline
		${\cal A}_{W^2H^2}(1_{W_-^a},2_{H_j},3_{W_-^b},4_{H^\dagger_i})$  &  $0$ & $\frac{1}{3}$ & 
		$\times \,C_{W^2H^2}\delta_{ab}\delta_{ij}$\\ 
		\hline
		$\mathcal{A}_{B^2H^2}(1_{B_-} ,2_{B_-} ,3_{H_i}, 4_{H_i^\dagger})$ &  1 & 0 & $\times\, C_{B^2H^2}$\\ 
		\hline
	\end{tabular}
	\caption{Values of the coefficients $a^{J}$ in the $s$-channel for the different  SM amplitudes (up to $J=1$) and  SM EFT amplitudes at $O(E^2/\Lambda^2)$ discussed in the text. For ${\cal A}^{I=0}_{\rm SM}$  we give the regularized coefficient $a^J|_{\rm reg}$ (see main text).}
	\label{tableSMEFT}
\end{table}

Let us now consider an example in the SM EFT in which IR divergences are present.
In particular, let us calculate  the self-renormalization  of
\be\label{B2H2} 
\mathcal{A}_{B^2H^2}(1_{B_-} ,2_{B_-} ,3_{H_i}, 4_{H_i^\dagger})\,=\frac{C_{B^2H^2}}{\Lambda^2}\la 12\ra^2\,.
\ee
We focus on the corrections proportional to $g_2$, which arise purely from the $s$-channel.
The amplitude \eq{B2H2}  only contributes with $J=0$  and  $SU(2)_L$ isospin $I=0$
in the $s$-channel. 
When studying the contributions to the self-renormalization  of \eq{B2H2},
we therefore just need to consider the amplitude
$\mathcal{A}_{\rm SM}(1_{H},2_{H^\dagger},3_{H^\dagger},4_{H})$
projected into the zero-isospin channel (due to isospin conservation). This reads
\be\label{4higgs}
\mathcal{A}^{I=0}_{\rm SM}(1_{H},2_{H^\dagger},3_{H^\dagger},4_{H})=
\frac{3}{4}\,g_2^2\left( \frac{1}{2}+\frac{u}{t} \right).
\ee
From \eq{tu},
we see that the last term of \eq{4higgs}, which comes from $t$-channel exchanges
of $W$s, has a $\theta\to 0$ singularity. 
Therefore, we have to use the regularized partial-wave coefficients defined in \eq{aIRsafe}.
${\bm T}_{\rm soft}$, once projected into the $I=0$-channel, is given by
\be\label{4Hsoft}
{\bm T}^{I=0}_{\rm soft}=
-\frac{3}{4}\,g_2^2\,.
\ee
Plugging Eqs.~\eqref{4higgs} and \eqref{4Hsoft} into
\eq{aIRsafe}, we get $a_{\rm SM}^{J=0}|_{\rm reg}=-3g_2^2/8$ and
 $a_{\rm SM}^{J\geq 1}|_{\rm reg}=2H_J{\bm T}^{I=0}_{\rm soft}$, with $H_J$ being the $J$-th harmonic number.
Using \eq{schannel} together with the collinear contributions in \eq{coll}, we obtain
 \be
 \gamma_{B^2H^2}=-\frac{1}{8\pi^2}\, a^{J=0}_{\rm SM}|_{\rm reg} \, C_{B^2H^2}+\gamma_{\rm coll} C_{B^2H^2}=
 -\frac{9}{64\pi^2}\,g_2^2 \, C_{B^2H^2}\,,
 \ee
 where we have also used that $\gamma_{\rm coll}=2\gamma_{\rm coll}^H=-3g_2^2/16\pi^2$
 \cite{EliasMiro:2020tdv,Jiang:2020mhe}.

We finish by commenting on certain cases in the SM EFT in which  the partial-wave decomposition is not so
 useful, since the number of terms in the sum in \eq{NMF} is infinite.
This   can occur in the calculation of the renormalization 
of  a 4-point amplitude from a 3-point amplitude.
An example can be found in \cite{us}: the renormalization of ${\cal A}_{WHle}$ from 
${\cal A}_{W^3}(1_{W^a_-},2_{W^b_-},3_{W^c_-})$,
given by the amplitudes  shown in Fig.~8 of \cite{us}. In this case, both $a^J_L$ and $a^J_R$  are nonzero for infinitely many $J$s,
a fact which is related to the presence of a logarithm in each channel \cite{us}.
When the infinite sums of the different channels are added, 
 only the contribution from a few $J$s remains nonzero, while the rest cancels. 
This is due to the fact that the logarithms cancel when adding all channels, leaving a constant term proportional to ${\cal A}_{WHle}$.
We therefore see that  in this particular case, \eq{NMF} is not very efficient to calculate the anomalous dimension.

\subsection{Nonlinear sigma models}\label{pions}

Let us now apply our methods to study the renormalization of nonlinear sigma models. Our analysis will allow us to easily obtain a very general analytic expression for the anomalous-dimension matrix. 

From the amplitude perspective, the model is defined as an EFT of real scalars, transforming under some symmetry group, and with amplitudes satisfying Adler's zero condition \cite{Adler:1964um}:
 the amplitudes vanish in the limit in which any of the external momenta is taken to zero. 
 For definiteness, we focus on scalars in the fundamental representation of $SO(N)$,
 and  only study 4-point amplitudes in an expansion in $E/\Lambda$.
Any  4-point amplitude involving  (positive) powers of the external momenta automatically satisfies Adler's zero condition.
Therefore, by starting at $O(E^2/\Lambda^2)$,  we can easily obtain  the 4-point amplitudes of this EFT  by just requiring that they are invariant under $SO(N)$. Higher-point amplitudes can  be constructed from these 4-point amplitudes by demanding proper factorization. As shown in \cite{Susskind:1970gf,Low:2019ynd}, imposing Adler's zero condition on the higher-point amplitudes then requires to introduce additional higher-point 
 contact interactions whose coefficients are completely fixed as a function of the Wilson coefficients $C_i$.
 In this sense the 4-point amplitudes play the role of ``building-blocks".\footnote{We note  that at $O(E^6/\Lambda^6)$  and higher,  $n$-point amplitudes with $n>4$ exist which are also ``building-blocks'' (having the property that they satisfy Adler's zero condition already by themselves) \cite{Low:2019ynd}. In the Lagrangian description these amplitudes correspond to operators whose leading contact interactions involve more than 4 fields. 
 One needs to extend our approach to higher-point amplitudes to cover these cases.}

Notice that we have only specified the unbroken group $SO(N)$ (and its representation) but not the broken group of the nonlinear sigma model. 
Remarkably, information about the coset structure can  be obtained by studying the double soft limit of the amplitudes, where two external momenta are taken to zero simultaneously \cite{ArkaniHamed:2008gz,Kampf:2013vha}. 
The relevant coset in our case is $SO(N+1)/SO(N)$.

Let us   find   the  building-block 4-point amplitudes for this EFT.
To this end, we first notice that any 4-point amplitude of $N$ real scalars transforming in the fundamental representation of  $SO(N)$ can be written as\footnote{Note that, for $SO(4)$, another flavor structure with the right transformation properties is $\epsilon_{ijkl}$, which gives rise to additional amplitudes. We will not consider this case further. }
\be
\label{PionAmplitudes}
{\cal A}(1^i,2^j,3^k,4^l) \, = \, f_s(t,u) \, \delta_{s} \, + \, f_t(u,s) \, \delta_{t}  \, + \, f_u(s,t) \, \delta_{u} \, ,
\ee
where $s,t,u$ 
are the Mandelstam variables and $i,j,k,l$ are flavor indices. The flavor-$SO(N)$ invariant tensors  read
\be
\delta_s = \delta_{ij} \delta_{kl} \,, \quad \delta_t = \delta_{ik} \delta_{jl}  \,, \quad \delta_u = \delta_{il} \delta_{jk} \, ,
\ee
and they transform under crossing like $s$, $t$ and $u$, respectively. Imposing crossing invariance, one finds that $f_s=f_t=f_u\equiv f$, and that $f$ is symmetric in its two arguments. We can then obtain the set of all independent amplitudes at order $w$ in $E/\Lambda$ 
by finding all linearly-independent functions $f(t,u)$
which are symmetric in $t$ and $u$ and are polynomials of degree $w/2$ in these variables ($w$ is always even
here). 
For later convenience, we choose the basis for these functions as\footnote{To see that this is a basis, notice that alternatively we could consider 
$f_{wr}(t,u) = t^{r/2} u^{(w-r)/2}+t^{(w-r)/2} u^{r/2}$ for the same range of $r$.
This covers all symmetric polynomials in $t$ and $u$ of degree $w/2$ in these variables. The number of these polynomials equals the number of polynomials in Eq.~\eqref{KinematicStructures}. Since Legendre polynomials are linearly independent,  we see that Eq.~\eqref{KinematicStructures}  forms indeed a basis. 
}
\be
\label{KinematicStructures}
f_{wr}(t,u) \, \equiv \, P_r\left(\frac{t-u}{s} \right) s^{w/2}\, \ \ \ (w=2,4,...)\,,
\ee
where $r=0, 2, 4, \dots, w/2$ $(r=0, 2, 4, \dots, w/2-1)$ if $w/2$ is even (odd). $P_r$ is the $r$-th Legendre polynomial.
We then find that the  building-block 4-point amplitudes for this EFT  are
\be
{\cal A}_{wr}(1^i,2^j,3^k,4^l)=\ \frac{C_{wr}}{F_\pi^{w}} \bigl( f_{wr}(t,u) \, \delta_{s} \, + \, f_{wr}(u,s) \, \delta_{t}  \, + \, f_{wr}(s,t) \, \delta_{u} \bigr)\, ,
\label{basis}
\ee
where $C_{wr}$ are Wilson coefficients. We have traded the scale $\Lambda$ for the pion decay constant $F_\pi$ which is defined by fixing the Wilson coefficient of the leading amplitude at $O(E^2)$ to one, {\it i.e.}~$C_{2\,0}=1$.

We will perform a decomposition into angular momentum in the $s$-channel, 
which will allow us to considerably simplify the computation.
In the same spirit,
it is also convenient to perform an ``isospin" decomposition of  the amplitude:
\be
{\cal A}_{wr}= \sum_{I=0}^2 {\cal A}_{wr}^{I} \, \Delta_I\,,
\label{Itb}
\ee
with flavor structures defined as 
\be
\label{IsospinflavorBasis}
\Delta_0 \equiv \frac{\delta_s}{N}\, , \quad \
\Delta_1 \equiv \frac{1}{2}\left( \delta_t - \delta_u \right)\, , \quad \
\Delta_2 \equiv \frac{1}{2}\left( \delta_t + \delta_u -\frac{2}{N} \delta_s \right) \, .
\ee
For $N=3$, projecting to one of these basis elements corresponds to
taking the initial (and final) state with definite isospin $I$.
For generic $N$, it amounts to choosing states in singlet, anti-symmetric
or traceless symmetric configurations of flavor, respectively.
The flavor basis is orthonormalized, in the sense that
\be\label{consISO}
\sum_{i'j'}(\Delta_{I})_{iji'j'}  (\Delta_{I'})_{i'j'kl} \, = \, \delta_{I I'} \, (\Delta_{I})_{ijkl} \, ,
\ee
where the orthogonality reflects the conservation of isospin. 

After the decomposition according to isospin, we next decompose the amplitudes into components with fixed angular momentum. Using Eq.~\eqref{pwd}, we can write
\be
\label{A_niRExpansion}
{\cal A}_{wr} \, = \, \left( \frac{s}{F_\pi^2} \right)^{w/2}   \sum_{IJ} n_J \,  P_J(c_\theta)  \, a^{IJ}_{wr} \, \Delta_{I} \, ,
\ee
where we have used that $d_{00}^J(\theta)=P_J(c_\theta)$ with  $P_J$ being the $J$-th Legendre polynomial.
Inverting this, the coefficients are given by
\be
\label{AIJformula}
a^{IJ}_{wr}=   \frac{1}{2} \left( \frac{s}{F_\pi^2}\right)^{-w/2} \hspace{-.1cm} \int_0^\pi d\theta \, s_\theta \,  P_J(c_\theta) \, {\cal A}_{wr}^I   \, .
\ee
Obtaining ${\cal A }^I_{wr}$ from  \eq{basis} and \eq{Itb}, and inserting it into \eq{AIJformula},  we get
\be
a^{IJ}_{wr}=   C_{wr} \left( 2  \, \kappa^J_{wr} \, + \, \frac{N}{n_J} \delta_{0I} \, \delta_{rJ} \right),
\ee
if $I,J$ are both even or both odd, and $a^{IJ}_{wr}=0$ otherwise. We have introduced 
\be
\kappa^J_{wr} \, \equiv \, \frac{(-1)^{w/2+J} \, [(w/2)!]^2}{(w/2-J)! \, (w/2+J+1)!} \  {}_4 F_3\left(-r, 1+r,-1-J-\frac{w}{2},J-\frac{w}{2}; 1,-\frac{w}{2},-\frac{w}{2}; 1 \right),
\ee
where $ {}_4 F_3$ is a generalized hypergeometric function.\footnote{This can alternatively be written as $\kappa^{J}_{wr} = \sum_{k=0}^r \frac{(-1)^{w/2+J-k}(r+k)! \, [(w/2-k)!]^2 }{[k!]^2 (r-k)! \, (w/2+J+1-k)! \, (w/2-J-k)!}$.}   We find that $a^{IJ}_{wr}=0$ for $J > w/2$. This means that only a finite number of angular-momentum states contribute in ${\cal A}_{wr}$.

We now consider the one-loop correction
involving amplitudes ${\cal A}_{w_Lr_L}$ on the left and ${\cal A}_{w_Rr_R}$ on the right of \eq{NMF}. Let us denote the resulting contribution to the anomalous dimension of the amplitude ${\cal A}_{wr}$ by $\Delta\gamma_r$. We suppress the dependence of $\Delta\gamma_r$ on $w_L, r_L,w_R,r_R$ to avoid clutter.  
Since there are no IR divergencies,
the total anomalous dimension $\gamma_r$  is  obtained by summing the different $\Delta\gamma_r$ for all $w_L$ and $w_R$ such that $w= w_L+w_R$ and for the ranges of $r_L$ and $r_R$ 
as given below Eq.~\eqref{KinematicStructures}.

The $s$-channel contribution to the anomalous dimensions follows from Eq.~\eqref{NMF}, {with a factor 1/2} to account for the case of identical particles in the internal legs (this factor is compensated by the sum over flavors if they are not identical).
The $t$- and $u$-channel contributions can be  obtained from this by crossing.
We then find that the anomalous dimensions  satisfy
\be
\label{PionLoopCorrection}
\sum_r \Delta\gamma_r \frac{{\cal A}_{wr}}{C_{wr}}\, = \, -\frac{1}{16\pi^2}\left( \frac{s}{F_\pi^2} \right)^{w/2}  \sum_{IJ}  n_J\, \Delta_I \, a_{w_L r_L}^{IJ} \, a_{w_R r_R}^{IJ}   \, P_J\left(\frac{t-u}{s}\right)  
+ \, (s \leftrightarrow t) \, + \, (s \leftrightarrow u) \, .
\ee
It is important to remark that the flavor structures $\Delta_I$ also change under crossing, as follows from Eq.~\eqref{IsospinflavorBasis}. Also  note that the LHS generically contains a
sum over all amplitudes of $O(E^{ w})$. 

In order to solve for the $\Delta\gamma_r$,
we next choose flavors such that $\delta_s=1, \delta_t=\delta_u=0$.
Then, we act with $\int_0^\pi d\theta s_\theta \, P_r(c_\theta)$ on both sides of  \eq{PionLoopCorrection}. This gives
\be
\label{AnomalousDimensionMatrix}
\Delta\gamma_r  =  -\frac{C_{w_L r_L} C_{w_R r_R}}{16 \pi^2 } \left(\frac{N}{n_r} \delta_{r_L r} \, \delta_{r_R r}  + 2 \, \delta_{r_L r} \, \kappa^r_{w_R r_R} + 2  \, \delta_{r_R r} \, \kappa^r_{w_L r_L} + 4 \hspace{-.4cm} \sum_{J=0}^{\smash{\min(\frac{w_L}{2},\frac{w_R}{2})}}   \hspace{-.4cm}  n_J n_r \,\kappa^J_{w_L r_L} \kappa^J_{w_R r_R} \kappa^r_{w J}\right),
\ee
where $n_r = 2r +1$. Notice that in the chosen basis, the dependence on $N$ only enters via the contributions  with $r_L=r_R=r$. The $\Delta\gamma_r$ therefore become very simple in the large-$N$ limit.

As an application of \eq{AnomalousDimensionMatrix}, 
we next present the anomalous dimensions for amplitudes up to $O(E^6)$. The two amplitudes at $O(E^4)$ are renormalized by the (single) amplitude at $O(E^2)$, corresponding to $w_L=w_R=2$ and $r_L=r_R=0$ in \eq{AnomalousDimensionMatrix}. This gives (recall that $C_{2\, 0}=1$)
\be
\gamma_0 \, = \, \frac{\frac{17}{9} -N}{16 \pi^2} \, , \quad \quad \,  \gamma_2 \, = \, -\frac{1}{72 \pi^2} \,.
\label{e2}
\ee
Similarly, the two amplitudes at $O(E^6)$ receive corrections from the product of an amplitude at $O(E^4)$ and the one at $O(E^2)$. 
The corresponding contributions to the anomalous dimensions in \eq{AnomalousDimensionMatrix} either have $w_L=2, w_R=4$ with $r_L=0$ and $r_R=0,2$, or $w_L=4, w_R=2$ with $r_L=0,2$ and $r_R=0$. Summing over all contributions we find
\be
\begin{gathered}
\gamma_0 \, = \, C_{4 \, 0} \, \frac{\frac{11}{36} -N}{8 \pi^2} \, - \, C_{4 \, 2} \, \frac{325}{288 \pi^2}\, , \\
\gamma_2\, = \, - C_{4 \, 0} \frac{5}{288 \pi^2} \, - \, C_{4 \, 2} \, \frac{65}{288 \pi^2} \, ,
\end{gathered}
\ee
where $C_{4\, 0}$ and $C_{4 \, 2}$ are the Wilson coefficients of the two amplitudes at $O(E^4)$. Using the fact that $\smash{SU(2) \times SU(2) / SU(2) \sim SO(4)/SO(3)}$, we can compare our results for the case $N=3$ with existing calculations for pions. Upon translating to our basis Eq.~\eqref{KinematicStructures},  we find that the anomalous dimensions of  \cite{Bijnens:1995yn} agree with the above results.

Let us also make an observation. Using Eqs.~\eqref{basis} and \eqref{e2}, we find that
\be
\sum_{r=0,2} \gamma_r \, \frac{{\cal A}_{2 \, r}}{C_{2 \, r}} \, = \, -\frac{1}{48 \pi^2  F_\pi^4} \left( (3 N -7) s^2+ 2  t^2 +2  u^2 \right) \delta_s \, + \, (s \leftrightarrow t) \, + \, (s \leftrightarrow u) \, .
\ee
For the case $N=3$, this has the interesting property that the only linear combination of ${\cal A}_{2\, 0}$ and ${\cal A}_{2\, 2}$ that is renormalized is crossing symmetric separately in kinematics and flavor, ${\cal A} \sim (s^2 +t^2+u^2) (\delta_s+\delta_t+\delta_u)$.\footnote{This was shown to also hold for any chiral $SU(N)$,
with flavor factor $\delta_s+\delta_t+\delta_u +N(d_s+d_t+d_u)/8$, where $d_s=d_{ijm}d_{klm}$,
$d_t=d_{ikm}d_{jlm}$ and $d_u=d_{ilm}d_{jkl}$, $d_{ijk}$ being the fully symmetric $SU(N)$ constants
\cite{talks}.}

Finally, let us comment on the connection to the Lagrangian description of the nonlinear sigma model with coset $SO(N+1)/SO(N)$.  The series of amplitudes necessary to satisfy Adler's zero condition discussed above are   equivalent to the series of contact interactions in the Lagrangian description that arise from expanding $SO(N+1)$-invariant operators in the number of fields \cite{Low:2019ynd}.  Each 4-point amplitude is thus in a one-to-one correspondence to an operator in the Lagrangian approach. Similarly, 
the anomalous-dimension matrix for the 4-point amplitudes that we have determined is equivalent to the anomalous-dimension matrix for the corresponding operators.

\subsection{Gravity}
\label{GRex}

We present here some examples of the use of \eq{NMF}
to calculate  anomalous dimensions in the EFT of spin-2 states.
Although these calculations are quite  lengthy with an ordinary  Feynman-diagrammatic  approach,
with the help of \eq{NMF} this is as easy as for the nonlinear sigma model
(or even simpler, since there are no complications with flavor).

Following the same approach as in the previous examples, the EFT of gravitons 
is defined by the  building-block
amplitudes obtained in an expansion  of $E/\Lambda$.
Einstein theory (General Relativity) corresponds to  the first possible amplitude  in this expansion.
This is a  3-point graviton interaction, which is fully determined by the little group:
\be
{\cal A}_{\rm GR}(1_{-},2_{-},3_{+})=\frac{1}{M_P}\left(\frac{\la 12\ra^3}{\la 13\ra \la 23\ra}\right)^2\,.
\label{GR}
\ee
From this, we can calculate higher-point amplitudes. 
For example,  the tree-level 4-point amplitude,   which we will  use later, is given by
\be
{\cal A}_{\rm GR_{+-}}\equiv {\cal A}_{\rm GR}(1_{+},2_{+},3_{-},4_{-})=\frac{1}{M_P^2}
\frac{[12]^4 \la 34\ra^4}{stu}\,.
\label{tree}
\ee
This can be  determined just from little-group scaling and by demanding proper factorization into 
the 3-point amplitudes in \eq{GR}.
 For the calculation of renormalization effects, we will later also need an amplitude with all gravitons having the same helicity.
No such amplitude can be constructed from \eq{GR} at tree-level. We must go to the one-loop level, where we have  \cite{Bern:2017puu}
\be
{\cal A}_{\rm GR_-}\equiv \,
{\cal A}_{\rm GR}(1_{-},2_{-},3_{-},4_{-}) \, =\, \frac{r}{16\pi^2M_P^4}\frac{\la 12\ra^4 \la 34\ra^4}{s^2}+{\rm crossing}
\, = \,\frac{r{\cal T}^2}{16\pi^2M_P^4} 
(s^2+t^2+u^2)\,,
\label{loop}
\ee
(and the corresponding complex conjugate), with $r=(N_F-N_B)/240$, $N_{F,B}$
being the number of fermions and bosons in the loop. Furthermore, we have defined
\be
{\cal T} \equiv \frac{\la 12\ra \la 34\ra} {[12][34]}\,.
\ee

At higher order in $E/\Lambda$, going beyond Einstein theory,  the first possible building-block  amplitude arises at order
$E^5/\Lambda^5$. It is the 3-point amplitude
\be
{\cal A}_{R^3}(1_{-},2_{-},3_{-})=\frac{C_{R^3}}{M_P^5}\la 12\ra^2\la 23\ra^2\la 31\ra^2\,,
\label{R3}
\ee
where we have traded $\Lambda$ for $M_P$ by redefining  $C_{R^3}$. We will do the same for the other amplitudes in the following.
In the field theory of gravity, \eq{R3}  arises from  the $R^3$ higher-dimensional term,
{\it i.e.}~a contraction of three Riemann tensors.
To get  new contact interactions, we have to go to  order  $E^8/\Lambda^8$.
Indeed, at this order we can have  two 4-point amplitudes:
\be
{\cal A}_{R^4}(1_{-},2_{-},3_{-},4_{-})=\frac{C_{R^4}}{M_P^8}\la 12\ra^4\la 34\ra^4+{\rm crossing}=\frac{C_{R^4}{\cal T}^2}{M_P^8}
(s^4+t^4+u^4)\,,
\label{R4}
\ee
and
\be
{\cal A}'_{R^4}(1_{-},2_{-},3_{+},4_{+})=\frac{C'_{R^4}}{M_P^8}\la 12\ra^4 [34]^4\,.
\label{R4b}
\ee

\begin{table}\centering
 \begin{tabular}{||c||c|c|c|c||} 
 \hline
& $J=0$ & $J=2$ &$J=4$ & \\
 \hline\hline
${\cal A}_{\rm GR_{+-}}$  & $0$ & $-6$ & $-\frac{25}{3}$ &  \\ 
 \hline
${\cal A}_{\rm GR_-}$  & $\frac{5}{3}$ & $\frac{1}{15}$ & 0& $\times\,\frac{r}{16\pi^2}$\\ 
 \hline\hline
$\widehat{\cal A}_{ R^3}$  & $\frac{1}{6}$ & $-\frac{1}{30}$ & 0 & $\times\, C_{R^3}$ \\ 
 \hline
${\cal A}_{R^4}$  &  $\frac{7}{5}$ & $\frac{4}{35}$ & $\frac{1}{315}$& $\times\, C_{R^4}$\\ 
 \hline
\end{tabular}
\caption{Values of the coefficients $a^{J}$ in the $s$-channel for the different 4-graviton amplitudes defined in the text.
For the tree-level GR amplitude ${\cal A}_{\rm GR_{+-}}={\cal A}_{\rm GR}(1_{+},2_{+},3_{-},4_{-})$ we give the regularized quantity defined in \eq{GRIRsafe} up to $J=4$.}
\label{tableGRAV}
\end{table}

We are now in a position to calculate the anomalous dimensions of the coefficients
$C_{R^3}$, $C_{R^4}$ and $C'_{R^4}$ at leading order.
We start with $C_{R^3}$. Its renormalization will be obtained from
the renormalization of the  4-graviton amplitude with equal helicities which arises 
 from the 3-graviton amplitudes in Eqs.~\eqref{GR} and \eqref{R3} at $O(E^6/\Lambda^6)$:
 \be
\widehat {\cal A}_{R^3}(1_{-},2_{-},3_{-},4_{-})=\frac{C_{R^3}}{M_P^6}{\cal T}^2 s tu\,.
\label{R34}
\ee
By simple dimensional analysis, 
one can realize that this amplitude  cannot be renormalized at the one-loop level,
since   no products  of tree-level amplitudes  can generate an amplitude of four gravitons
 of equal helicity at order $E^6/\Lambda^6$.
Indeed, it was shown in \cite{Bern:2017puu} that the leading nonzero contribution to the renormalization of  \eq{R34}
arises from  the product of the one-loop amplitude in \eq{loop} and the amplitude in \eq{tree}.
The calculation has IR divergencies which must be taken into account, as explained
in Sec.~\ref{soft}, by using the regularized coefficients \eq{aIRsafe}.
From \eq{NMF} we then get\footnote{Notice that the {statistical factor 1/2} for identical particles
in the internal lines  is compensated in \eq{R3cont} by the factor $2$ coming from the  two equal contributions
 $\int {\cal A}_{\rm GR}(1_{-},2_{-},\bar{2}'_{-},\bar{1}'_{-}){\cal A}_{\rm GR}(1'_{+},2'_{+},3_{-},4_{-})$ and
 $\int {\cal A}_{\rm GR}(1_{-},2_{-},\bar{2}'_{+},\bar{1}'_{+}){\cal A}_{\rm GR}(1'_{-},2'_{-},3_{-},4_{-})$.} 
\be
\gamma_{R^3}\widehat {\cal A}_{R^3}
= -\frac{C_{R^3}}{8\pi^2}\left(\frac{s}{M_P^2}\right)^3
\sum_J n_J\, a^{J}_{\rm GR_-} a^{J}_{\rm GR_{+-}}|_{\rm reg}\, P_J\left(\frac{t-u}{s}\right)+{\rm crossing}\,,
\label{R3cont}
\ee
where the coefficients $a^{J}_{\rm GR_{+-}}|_{\rm reg}$  are defined
in \eq{aIRsafe}, with ${\bm T}_{\rm soft}={-2s}/{M_P^2}$ and the replacement $\int s_{\theta/2}^{-2}\to 2\int s_{\theta}^{-2}$
due to the identical particles in the internal lines. Using \eq{tree}, we find
\be
a^{J}_{\rm GR_{+-}}|_{\rm reg}
=-4 H_J\,,
\label{GRIRsafe}
\ee
where $H_J$ is the $J$-th harmonic number. The values of the coefficients $a^{J}_{\rm GR_-}$ and
 $a^{J}_{\rm GR_{+-}}|_{\rm reg}$ are given in Table~\ref{tableGRAV}, where we see that
they are simultaneously nonzero  only for $J=2$. We then get
\be
\gamma_{R^3}\widehat {\cal A}_{R^3}
= \frac{C_{R^3}}{4\pi^2}\frac{r}{16\pi^2}\frac{s^3}{M_P^6}
P_2\left(\frac{t-u}{s}\right)+{\rm crossing}\,.
\label{r3p}
\ee
One can check that adding the crossed terms in \eq{r3p} makes the RHS
proportional to $\widehat {\cal A}_{R^3}$, as it should be.
Nevertheless,  since we are only interested in the value of $\gamma_{R^3}$,
it is simpler  to project both sides of   \eq{r3p}
 into  some specific kinematics, e.g.~$t= u=-s/2$.
This gives
\be
\gamma_{R^3}=\frac{r}{16\pi^4}\left( P_2(0)-\frac{1}{4} P_2(3)\right)=
-\frac{60r}{(4\pi)^4}\,.
\ee
Although this  result was  already obtained in  \cite{Bern:2017puu} by on-shell methods,
our formula allows us to understand the dependence  
of \eq{r3p} on the Mandelstam variables: this
is indeed determined by the fact that, in the $s$-channel, only  internal states with
$J=2$  contribute to $\gamma_{R^3}$.

Recycling the above result, we can easily obtain
the anomalous dimension of $C_{R^4}$. It can again only  arise from the partial waves 
of   $\widehat {\cal A}_{R^3}$ and ${\cal A}_{\rm GR_{+-}}$ with $J=2$, leading to
\be
\gamma_{R^4}{\cal A}_{R^4}=
-C_{R^4} \frac{5}{8\pi^2}\left(\frac{s}{M_P^2}\right)^4
 a^{J=2}_{R^3}\, a^{J=2}_{\rm GR_{+-}}|_{\rm reg}\, P_2\left(\frac{t-u}{s}\right)+{\rm crossing}\,.
\ee
Using Table~\ref{tableGRAV}, this gives
\be
\gamma_{R^4}=-\frac{{C_{R^3}}}{8\pi^2}\,.
\ee
At the one-loop level, we do not find any  contribution to the anomalous
dimension of $C'_{R^4}$, due to  the helicities in ${\cal A}'_{R^4}$.

\section{Conclusions}

We have here exploited the power of  angular-momentum analysis
to reduce the computation of  one-loop anomalous dimensions
to a sum of products of partial-wave coefficients, \eq{NMF}.
For the anomalous dimensions of contact interactions (higher-order amplitudes in EFTs),
the sum reduces to a few  terms,  making the calculation quite straightforward.
 We have also shown that \eq{NMF} remains valid in the presence of IR divergencies, once the partial-wave coefficients are regularized according to \eq{aIRsafe}.

The classification of the possible angular momenta $J$ contributing to the renormalization of the EFT amplitudes ${\cal A}_i$
has turned out to be useful since it tells us about the origin of the
anomalous dimensions $\gamma_i$, possible selection rules, 
and  potential relations between different $\gamma_{i}$,
not only inside the same theory but also between different theories.
In this sense, the angular-momentum analysis has provided a rational for the ``universality" of some anomalous dimensions,
hinted at in \cite{us,Jiang:2020mhe},
which  remains hidden  when performing calculations with ordinary Feynman diagrams.

We have shown this explicitly in  several examples  
for the SM EFT, where 
a class of one-loop mixings were  found to be proportional to  the same 
 coefficient $a_{\rm SM}^J$ (see \eq{mixing}).
We have also analyzed  the renormalization of nonlinear sigma models,
and shown how to use \eq{NMF}  to easily calculate the 
anomalous dimensions of  4-point amplitudes at any order in $E/\Lambda$.
As a last example, we have applied \eq{NMF} to obtain   anomalous dimensions  in  the EFT of gravity,
where the simplicity of the on-shell method has no competitor.
In particular, we have seen that 
only the $J=2$ partial wave contributes to $\gamma_{R^3}$, explaining 
the dependence  of \eq{r3p} on the  Mandelstam variables found in \cite{Bern:2017puu}.

Similarly as for the angular momentum, 
it can also be  useful to   decompose the amplitudes according to other conserved quantum numbers. 
For the case of  isospin, we have already seen an example in the SM EFT in the renormalization of \eq{B2H2},
and another one in nonlinear sigma models. This decomposition leads also to very useful selection rules, as 
the amplitudes ${\cal A}_{L,R}$ must have the same isospin as the one that they renormalize, ${\cal A}_i$.

We have focused here on the one-loop renormalization of 4-point amplitudes.
But we do not see any obstacle to the extension of this method to higher-point
amplitudes or higher-loop order, since we have mainly  relied on  angular-momentum conservation
and   the fact that the anomalous dimension can be obtained from the product of  two amplitudes $\gamma{\cal A}\sim\int d{\rm LIPS}\, {\cal A}_L\,{\cal A}_R$,  obtained from cutting loop diagrams.
These extensions are left for the future.

\section*{Acknowledgments}
P.B. thanks Max Ruhdorfer and Elena Venturini for useful discussions, has been partially supported by the DFG Cluster of Excellence 2094 ORIGINS, the Collaborative Research Center SFB1258, and thanks the Munich Institute for Astro- and Particle Physics (MIAPP) for hospitality, which is funded by the Deutsche Forschungsgemeinschaft (DFG, German Research Foundation) under Germany's Excellence 
Strategy - EXC-2094 - 390783311. C.F. is supported  by the fellowship FPU18/04733 from the Spanish Ministry of Science, Innovation and Universities. A.P. is supported by the Catalan ICREA Academia Program and  grants FPA2017-88915-P, 2017-SGR-1069 and SEV-2016-0588.

\appendix

\section{IR divergencies in the Passarino-Veltman\\decomposition}
\label{appA}

The purpose of this Appendix is to prove the validity of \eq{MF} for computing anomalous dimensions, corrected with Eqs.~\eqref{IR} and \eqref{coll} in the presence of IR divergencies. The strategy consists in using the Passarino-Veltman (PV)
decomposition.
In particular, we will discuss the structure of
IR divergencies of one-loop amplitudes from the PV point of view.
For simplicity, we will only  consider 4-point amplitudes.

A one-loop  amplitude can be decomposed in the PV basis in the following way:
\be\label{PVexp}
\mathcal{A}_{\rm loop}=\sum_a C_2^{(a)}I_2^{(a)}+\sum_b C_3^{(b)}I_3^{(b)}+\sum_c C_4^{(c)}I_4^{(c)}+R\,,
\ee
where the scalar integrals $I_{2,3,4}$, corresponding to bubbles, triangles
and boxes respectively, are defined e.g.~in \cite{us}, and $R$ is a rational term.
In general, $\mathcal{A}_{\rm loop}$ can contain IR divergencies,
which we want to subtract to obtain an IR-finite result.
The IR-divergent part of $\mathcal{A}_{\rm loop}$,
which is well-known for gauge theories and gravity \cite{Catani:1998bh,Becher:2009cu,Dunbar:1995ed},
has the following structure:
\be\label{PVIR}
\mathcal{A}_{\rm loop}^{\rm IR}=\sum_a \hat{C}_2^{(a)}I_2^{(a)}+\sum_b \hat{C}_3^{(b)}I_3^{(b)}\,.
\ee
The difference
\be\label{PVreg}
\mathcal{A}_{\rm loop}-\mathcal{A}_{\rm loop}^{\rm IR}=
\sum_a \left[{C}_2^{(a)}-\hat{C}_2^{(a)}\right]I_2^{(a)}+\sum_b \left[{C}_3^{(b)}-\hat{C}_3^{(b)}\right]
I_3^{(b)}+\sum_c C_4^{(c)}I_4^{(c)}+R\,,
\ee
is then guaranteed to be IR safe. To compute anomalous dimensions $\gamma$ at the one-loop order,
we need to extract the coefficient of the $\epsilon^{-1}$ UV-divergent part,
which is simply proportional to the sum over bubble coefficients:
\be
\gamma\mathcal{A}=-2 \,\mathcal{A}_{\rm loop}^{\rm UV}=-\frac{1}{8\pi^2}\sum_a \left[{C}_2^{(a)}-\hat{C}_2^{(a)}\right].
\ee
For this purpose we follow the generalized unitarity method, which consists in performing 2-cuts
of \eq{PVreg}:\footnote{We normalize the 2-cut as ${\rm cut}[I_2]=-1/8\pi^2$ to make
\eq{MF} equal to the sum over the 2-cuts of $\mathcal{A}_{\rm loop}$.}
\be
{\rm cut}^{(a)}\left[\mathcal{A}_{\rm loop}-\mathcal{A}_{\rm loop}^{\rm IR} \right]=
-\frac{C_2^{(a)}-\hat{C}_2^{(a)}}{8\pi^2}\,+{\rm cut}^{(a)}\left[\sum_b C_3^{(b)}|_{\rm reg}\,I^{(b)}_3+\sum_c C^{(c)}_4I^{(c)}_4 \right] ,
\ee
where $C_3|_{\rm reg}=C_3-\hat{C}_3$.
The terms involving 2-cuts of triangles and boxes are in general different from zero,
but it was shown in \cite{us} that, for IR-safe amplitudes like \eq{PVreg},
they cancel in the sum over all 2-cuts.\footnote{The proof in \cite{us} can be easily readapted
by substituting $C_3\to C_3|_{\rm reg}$.} We then obtain, by summing over all possible 2-cuts,
\be\label{proof}
\sum_a {\rm cut}^{(a)}\left[\mathcal{A}_{\rm loop}-\mathcal{A}_{\rm loop}^{\rm IR} \right]
=\gamma\mathcal{A}\,.
\ee
In the absence of IR divergencies ($\mathcal{A}_{\rm loop}^{\rm IR}=0$), \eq{proof}  matches with \eq{MF}.

\begin{figure}[t]
\centering
\includegraphics[width=.45\textwidth]{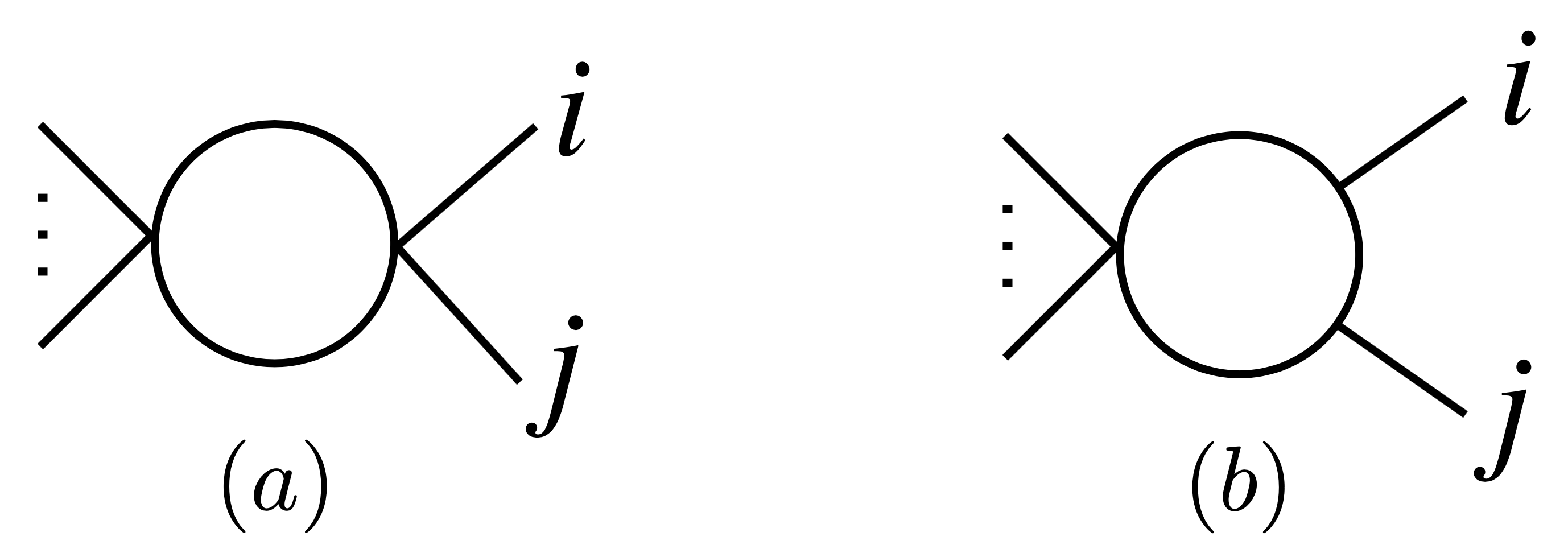}
\caption{\it Bubble (a) and triangle (b) topologies which appear in the PV decomposition
of $\mathcal{A}_{\rm loop}^{\rm IR}$
in gauge and gravity theories.}
\label{topologies}
\end{figure}

 To treat the case in which IR divergencies are present,  we need to know the  explicit form  of
$\mathcal{A}_{\rm loop}^{\rm IR}$. 
For definiteness, we first consider gauge theories, and comment later on the case of gravity. 
By expanding the one-loop expressions
given in \cite{Catani:1998bh,Becher:2009cu} on the PV basis, we find  nonzero bubble coefficients related to
collinear IR divergencies:
\be\label{bubbles}
\hat{C}_2^{(ij)}=-8\pi^2 \, \gamma_{\rm coll}^{(i)}\,
\frac{{\bm T}_i \cdot{\bm T}_j}{{\bm T}_i^2}{\cal A}_{\rm tree}+(i\leftrightarrow j)\,,
\ee
where the external  legs $i$ and $j$ are as shown in  Fig.~\ref{topologies} ($a$),
and we follow the color-space notation usually adopted,
with color/flavor indices left implicit \cite{Catani:1998bh,Becher:2009cu}. 
When we unfold them, we have
\be
 {\cal A}_{\rm tree}\to{\cal A}_{\rm tree}^{abcd}\ \ {\rm and}\ \  {\bm T}_i\to (T_i^A)_a^{a'}\,,
 \ee
 where the generator $T_i$ is in the representation appropriate
for particle $i$. The dot-product is ${\bm T}_i \cdot{\bm T}_j\equiv{\bm T}_i^A{\bm T}_j^A$ (see e.g.~\cite{Catani:1998bh} for more details).
We then sum over all ${\rm cut}^{(ij)}$ with $i<j$, where the indices $i,j$ label the two legs that are cut out from the rest of the states. The bubble contributions then reduce to a sum over particle legs:
\be
-\sum_{i<j}{\rm cut}^{(ij)}\left[\mathcal{A}_{\rm loop}^{\rm IR}\right]|_{\rm bubble}=\frac{1}{8\pi^2}\sum_{i<j}\hat{C}_2^{(ij)}=\sum_i \gamma_{\rm coll}^{(i)}{\cal A}_{\rm tree}=\gamma_{\rm coll}{\cal A}_{\rm tree}\,,
\ee
thanks to the property of color/flavor conservation, $\sum_j {\bm T}_j {\cal A}_{\rm tree}=0$. This reproduces \eq{coll}. 

On the other hand, soft IR divergencies give contributions  to 1-mass triangles,
with coefficients
\be
\hat{C}_3^{(ij)}=-g^2 s_{ij} \,{\bm T}_i \cdot{\bm T}_j\,{\cal A}_{\rm tree}\,,
\ee
where $i$ and $j$ label the external  legs, as shown in Fig.~\ref{topologies} ($b$).
In this case,  a 2-cut gives
\be
{\rm cut}^{(ij)}\left[\mathcal{A}_{\rm loop}^{\rm IR}\right]|_{\rm triangle}=\hat{C}_3^{(ij)}{\rm cut}[I_3^{(ij)}]
=\frac{1}{4\pi^3}\,g^2 \,{\bm T}_i \cdot{\bm T}_j\,{\cal A}_{\rm tree}\int d{\rm LIPS}\,\frac{1}{s_{\theta'\!/2}^2}\,.
\label{cutir3}
\ee
 In the last step, we could have equally well expressed the cut of the triangle as $\int {c_{\theta'\!/2}^{-2}}$,
which is the same after the substitution $\theta' \to (\pi-\theta')$,
or as $2 \int {s_{\theta'}^{-2}}$, obtained by symmetrization over the interval $[0,\pi]$.
We will use \eq{cutir3} when either ${\cal A}_L$ or ${\cal A}_R$ is singular for $\theta\to 0$.
In this case  we have ${\cal A}_{L,R}\sim s^{-2}_{\theta/2}$, 
and  the integrand in \eq{aIRsafe}   is well-behaved.
When  ${\cal A}_{L,R}\sim s^{-2}_{\theta}$, as it happens when the two incoming/outgoing particles are identical,
 we need to replace $\int s^{-2}_{\theta'\!/2}$  in \eq{cutir3} by $2 \int {s_{\theta'}^{-2}}$, and similarly for \eq{aIRsafe}.

Using \eq{cutir3} in \eq{proof}, we obtain \eq{IR}, with the identification
${\bm T}_{\rm soft}^{ij}=g^2 \,{\bm T}_i \cdot{\bm T}_j$. 
Notice that in QED we simply have ${\bm T}_{\rm soft}^{ij}=e^2 q_i q_j$.
For gravity, IR divergencies have the same PV structure as for gauge theories, 
with $\gamma_{\rm coll}=0$ and ${\bm T}_{\rm soft}^{ij}=-2 s_{ij}/M_P^2$
\cite{Dunbar:1995ed}.

{Soft singularities are also present in
scalar amplitudes involving the exchange of a massless scalar.
The methods discussed here and in Sec.~\ref{soft} can be easily extended to this case.
Instead, 4-point amplitudes involving Yukawa interactions never feature
$\theta\to 0$ singularities, as they go at worst like $\theta^{-1}$ (and the same
for $\theta\to \pi$).}

\section{Partial-wave decomposition}
\label{AppB}

Let us consider the  process 
$1,2\to 3,4$, 
where  the incoming pair $1,2$ lies in the direction
defined by the polar angles $(\psi,\omega)$ and the outgoing pair $3,4$
 in the direction defined by the  angles $(\theta,\phi)$.
 By using  the completeness of the angular momentum basis, the amplitude can be written as
\bea
\label{pwexp}
&&{\cal A}(s,\theta,\phi;\psi,\omega)\equiv 
\la \theta\phi; h_3h_4| \mathcal{T}|\psi\omega;h_1h_2\ra
\nonumber\\
&=&\sum_{JJ'MM'}\la\theta\phi ; h_3h_4 | J'M';h_3 h_4\ra \la J'M' ; h_3h_4 | \mathcal{T} | JM;h_1h_2\ra \la JM ; h_1h_2| \psi\omega;h_1h_2\ra\,.
\eea
Due to  angular momentum conservation, we have
\be\label{jconse}
\la J'M' ; h_3h_4| \mathcal{T} | JM;h_1h_2\ra=  \delta_{MM'} \, \delta_{JJ'}\left(\frac{\sqrt{s}}{\Lambda}\right)^w a^J.
\ee
Applied to \eq{pwexp}, this  gives
\bea
{\cal A}(s,\theta,\phi;\psi,\omega)=
\left(\frac{\sqrt{s}}{\Lambda}\right)^w
 \sum_{JM}n_J\, e^{i\phi(M-h_{34})}d_{Mh_{34}}^J(\theta)\,e^{-i\omega(M-h_{12})}d_{Mh_{12}}^J(\psi)\, a^J ,
\label{general}
\eea
where we have defined $n_J=2J+1$ and have introduced the Wigner $d$-functions
\be\label{jbasis}
{d}^J_{Mh_{12}}(\theta)=\frac{e^{i\phi(M-h_{12})}}{\sqrt{n_J}}\la JM;h_1 h_2 | \theta\phi; h_1 h_2\ra \, .
\ee
They can be expressed in terms of trigonometric functions as 
{\begin{multline}
{d}^J_{MM'}(\theta)= \Bigl[(J+M)! \, (J-M)! \, (J+M')! \, (J-M')!  \Bigr]^{1/2}\\ 
\times \sum_S \left[\frac{(-1)^{M'-M+S} \, c_{\theta/2}^{~2J+M-M'-2S} \, s_{\theta/2}^{~M'-M+2S}}{(J+M-S)! \, S! \, (M'-M+S)! \, (J-M'-S)!}\right] ,
\end{multline}}

\noindent where the sum is over all $S$ such that the arguments of the factorials in the denominator are nonnegative.
The Wigner $d$-functions fulfil an orthogonality condition given by
\be
\int_0^\pi d\theta\, s_\theta \,d_{MM'}^J(\theta)d_{MM'}^{J'}(\theta)=\frac{2\delta_{JJ'}}{n_J} .
\label{ortho}
\ee
We always have the freedom to choose a frame such that $\psi=\omega=0$, giving
\be
{\cal A}(s,\theta,\phi)
=e^{i\phi(h_{12}-h_{34})}
\left(\frac{\sqrt{s}}{\Lambda}\right)^w
\sum_J n_J\, d_{h_{12}h_{34}}^J(\theta)\, a^J ,
\ee
where we have used the property $d^J_{MM'}(0)=\delta_{MM'}$.

\end{document}